\documentclass[final,3p,times,twocolumn,authoryear]{elsarticle}

\usepackage{amssymb}
\usepackage{amsmath}

\usepackage{multirow}
\usepackage{booktabs}
\usepackage{balance}

\usepackage[colorlinks,
linkcolor=blue,
anchorcolor=blue,
citecolor=blue]{hyperref}

\journal{}

\begin{document}

\begin{frontmatter}



\title{Extracting Inter-Protein Interactions Via Multitasking Graph Structure Learning}

\author{Jiang Li\corref{cor1}\fnref{label1}}
\affiliation[label1]{organization={School of Artificial Intelligence and Automation, Huazhong University of Science and Technology},
            city={Wuhan},
            postcode={430074},
            country={China}}
\ead{lijfrank@hust.edu.cn}
\cortext[cor1]{Corresponding author.}

\author{Yuan-Ting Li\fnref{label2}} 
\affiliation[label2]{organization={College of Life Sciences, South-Central Minzu University}, 
            city={Wuhan},
            postcode={430074}, 
            country={China}}
\ead{lyting1212@outlook.com}
\fntext[]{Received 18 January 2025; Received in revised form 18 January 2025; Accepted 18 January 2025}
\begin{abstract}
Identifying protein-protein interactions (PPI) is crucial for gaining in-depth insights into numerous biological processes within cells and holds significant guiding value in areas such as drug development and disease treatment. Currently, most PPI prediction methods focus primarily on the study of protein sequences, neglecting the critical role of the internal structure of proteins. This paper proposes a novel PPI prediction method named MgslaPPI, which utilizes graph attention to mine protein structural information and enhances the expressive power of the protein encoder through multitask learning strategy. Specifically, we decompose the end-to-end PPI prediction process into two stages: amino acid residue reconstruction (A2RR) and protein interaction prediction (PIP). In the A2RR stage, we employ a graph attention-based residue reconstruction method to explore the internal relationships and features of proteins. In the PIP stage, in addition to the basic interaction prediction task, we introduce two auxiliary tasks, i.e., protein feature reconstruction (PFR) and masked interaction prediction (MIP). The PFR task aims to reconstruct the representation of proteins in the PIP stage, while the MIP task uses partially masked protein features for PPI prediction, with both working in concert to prompt MgslaPPI to capture more useful information. Experimental results demonstrate that MgslaPPI significantly outperforms existing state-of-the-art methods under various data partitioning schemes.
\end{abstract}




\end{frontmatter}


\section{Introduction}
\label{sec:Introduction}
Protein-protein interactions (PPIs) play a crucial role in numerous biological processes within cells, including cell proliferation, metabolic cycles, DNA replication and transcription, immune responses, and signal transduction~\cite[]{Huang2020SkipGNN}. Accurate prediction of PPI not only helps deepen the understanding of protein functions but is also of great significance to fields such as medical diagnosis, disease treatment, and drug design~\cite[]{Raman2010Construction,Liu2021HPODNets}. Traditionally, PPI research has primarily relied on experimental techniques such as yeast two-hybrid~\cite[]{Fields1994The}, mass spectrometry protein complex identification~\cite[]{Ho2002Systematic}, crystallography and protein chips~\cite[]{Zhu2001Global}. However, these experimental methods have several limitations, including being time-consuming, costly, having a high false-positive rate, and being operationally complex~\cite[]{Uetz2000Acomprehensive,Luo2015AHighly}. Therefore, the development of computational methods with high accuracy and efficiency to precisely identify PPI types has become particularly urgent.

Some machine learning methods have been developed by researchers to compensate for the shortcomings of wet experiments. Early studies utilized methods such as support vector machines (SVM)~\cite[]{Chatterjee2011PPI_SVM} and random forests (RF)~\cite[]{You2015Predicting} for protein interaction prediction. Guo et al.~\cite[]{Guo2008Using} proposed a sequence-based computational method that combines a new feature representation of autocovariance and SVM to predict PPI data in Saccharomyces cerevisiae. Wang et al.~\cite[]{Wang2018Animproved} introduced a method based on the feature-weighted rotation forest algorithm for predicting PPI. Traditional machine learning methods largely depend on manually extracted protein sequence or structural features, which have inherent limitations in accurately extracting deep features of PPI. In recent years, deep learning technologies have been widely applied in the field of bioinformatics. Hashemifar et al.~\cite[]{10.1093/bioinformatics/bty573} proposed a model based on convolutional neural networks (CNN), combined with random projection and data augmentation techniques to predict PPI types. Chen et al.~\cite[]{10.1093/bioinformatics/btz328} introduced an end-to-end framework that captures the mutual influence of protein pairs by integrating deep residual recursive CNNs. ADH-PPI~\cite[]{asim2022adhppi} combines long short-term memory (LSTM), convolution, and attention mechanism to discover the most discriminative features and their short-term and long-term dependencies. Nambiar et al.~\cite[]{nambiar2020transforming} utilized Transformers to extract high-level features from amino acid sequences to enhance the performance of PPI prediction tasks. However, the primary or secondary structure of proteins is crucial for understanding their functions and interactions, and relying solely on protein sequences for PPI prediction still has certain limitations.

Methods based on protein structure generally employ graph neural networks (GNNs) to model the two-dimensional or three-dimensional structures of proteins, thereby enhancing the performance of PPI prediction to some extent. Gao et al.~\cite[]{gao2023hierarchical} constructed a hierarchical graph using PPI data and captured the structure-function relationship of proteins through graph learning models to enhance PPI prediction performance. Yuan et al.~\cite[]{yuan2021structure} regarded a protein as an undirected graph, integrated evolutionary and structural information as features of amino acid residues, and transformed PPI site prediction into a node classification problem. However, these methods require modeling the internal structure of each protein, thus consuming a significant amount of computational resources when aggregating information from amino acid residues. To address the aforementioned challenges regarding memory resources, MAPE-PPI~\cite[]{wu2024mapeppi} decoupled the PPI prediction task into two stages. This method first learns a microenvironment codebook based on amino acid residues and then combines GNN to extract higher-level protein features based on this codebook. However, MPAE-PPI still has three issues: (1) GNN achieved microenvironment functionality to some extent by aggregating information from neighbors, and training the microenvironment codebook consumed additional hardware resources, adding to the model's training burden; (2) In the protein graph, constructing a complex large graph by considering multiple types of edges (such as radius edge and K-nearest edge) led to information redundancy and increased waste of computational resources; (3) There was no learning of different weights for the importance of different residues, which led to inaccurate modeling of proteins.

To address the above challenges, in this work, we propose a multitask graph structure learning approach for protein-protein interaction prediction (MgslaPPI). The proposed MgslaPPI decomposes the end-to-end PPI prediction into two stages: amino acid residue reconstruction (A2RR) and protein interaction prediction (PIP), utilizing GNNs to respectively extract the internal structural information and external interaction information of proteins. In the A2RR stage, we employ graph attention network (GAT)~\cite[]{veličković2018graph} to model proteins. The input for this stage is a protein graph, where nodes represent amino acid residues and edges represent connections between residues. The goal of the A2RR stage is to capture the dependencies among amino acid residues, thereby mining the internal structural information of proteins. In the PIP stage, we use graph convolutional network (GCN)~\cite[]{pmlr-v119-chen20v} as the encoder to model the PPI network, thereby extracting PPI information between proteins. In this stage, we construct a PPI graph, where proteins serve as nodes and interactions between pairs of proteins serve as edges. Additionally, to enhance the expressive power of the graph encoder, we introduce two auxiliary tasks in the PIP stage, i.e., protein feature reconstruction (PFR) and masked interaction prediction (MIP). The PFR task uses a graph decoder to reconstruct the features of encoded proteins and narrows the gap between original and reconstructed features; the MIP task randomly masks parts of the protein features and uses a graph encoder with shared parameters to mine interaction information. These two auxiliary tasks aim to increase the difficulty of protein encoding by the graph encoder, thereby enhancing the expression of protein information. Through extensive validation experiments on two public protein interaction datasets, the results show that the proposed MgslaPPI method outperforms existing advanced PPI prediction methods.

The main contributions of this paper are as follows:
\begin{itemize}
     \item We propose a novel multitask graph structure approach called MgslaPPI for PPI prediction. This method captures the internal structural information and external interaction information of proteins through two stages, further enriching the research on structure-based PPI prediction methods.
     \item MgslaPPI employs a graph attention mechanism in the residue reconstruction stage to assign different weights to the contributions of different residues, which helps to accurately obtain the structural features of proteins. Moreover, this simple residue reconstruction method achieves low memory consumption and reduces the training burden of the model.
     \item In the interaction prediction stage, we integrate protein reconstruction and masked interaction prediction tasks, compelling the graph encoder to learn more useful information and thereby enhancing its protein expression capability.
     \item The results of numerous comparative and ablation experiments demonstrate that our MgslaPPI outperforms the current state-of-the-art baseline models under various data partitioning schemes.
\end{itemize}

The structure of this paper is organized as follows. Section~\ref{sec:Related_Work} reviews the related efforts in PPI prediction; Section~\ref{sec:Methodology} details the proposed method, MgslaPPI; Section~\ref{sec:Experiment} reports and discusses the experimental results; and the final Section summarizes the work.

\section{Related Work}
\label{sec:Related_Work}
PPIs are the core driving force behind many biological processes, regulating a multitude of cellular functions~\cite[]{Keskin2008Characterization}. The complex types of proteins and their interconnections within PPI networks contain key information about cellular processes and disease mechanisms~\cite[]{Hakes2008Protein,Wang2022Protein}. In-depth research on PPI can help human understand cellular functions and promote drug discovery. In the past, some machine learning techniques~\cite[]{Chatterjee2011PPI_SVM,You2015Predicting,Guo2008Using,Zhou2017Multi-scale,10.1002/prot.26486} were developed to address the challenges faced by traditional experimental methods. However, these shallow models, due to their simple network structures, were unable to capture highly nonlinear patterns and were overly reliant on manually extracted protein features. Recently, deep learning technologies, with their powerful feature learning capabilities, have gained widespread attention in the field of bioinformatics. Methods based on deep learning~\cite[]{Sun2017Sequence-based,Li2022SDNN-PPI,10.1093/bioinformatics/bty573,10.1093/bioinformatics/btz328,asim2022adhppi,dutta-saha-2020-amalgamation} have been emerging, mining complex relationships and information within PPI networks through multilayer nonlinear transformations. Hashemifar et al.~\cite[]{10.1093/bioinformatics/bty573} proposed a sequence-based CNN model that combines random projection and data augmentation techniques to capture complex and nonlinear relationships involved in protein interactions. Li et al.~\cite[]{molecules23081923} proposed a deep neural network framework called DNN-PPI, which automatically learns features of protein primary sequences through encoding, embedding, CNN, and LSTM, and then makes predictions using fully connected neural networks, demonstrating significant generalization capabilities.

Recent studies have considered protein correlations and utilized the graph structure to model PPI data. Huang et al.~\cite[]{Huang2020SkipGNN} proposed a model named SkipGNN, which aggregates information from direct and second-order interactions to predict molecular interactions, optimizing the GNN model using skip graphs and original graphs. Lv et al.~\cite[]{ijcai2021p506} introduced a GNN-based PPI prediction method that uses graph convolution to automatically learn protein features in PPI networks. Gao et al.~\cite[]{gao2023hierarchical} proposed a dual-view hierarchical graph learning model, consisting of bottom-layer and top-layer GNN representation learning, which can more effectively simulate the natural hierarchical structure of PPI. Zeng et al.~\cite[]{Zeng2024GNNGL-PPI} used graph isomorphism network (GIN)~\cite[]{xu2018how} to extract global graph features from PPI network graphs and GIN As Kernel to extract local subgraph features from subgraphs of protein vertices. Kang et al.~\cite[]{10.1093/bioinformatics/btad052} introduced ESM-1b encoding as feature input for protein sequences, integrating attention-free Transformer modules with graph attention network frameworks to fully express protein sequence features. Wu et al.~\cite[]{wu2024mapeppi} used heterogeneous graph neural networks to encode protein graphs, then quantized the microenvironment into discrete codes through vector quantization, while capturing dependencies between microenvironments through masked codes. In these methods, sequence-based models struggle to extract the internal structure of proteins, limiting the performance of PPI prediction; structure-based models are less studied and have significant resource overhead issues. To address the shortcomings of previous models, this study proposes a novel PPI prediction method named MgslaPPI. This method not only reduces memory overhead but also captures both the internal structural information and external interaction information of proteins simultaneously.

\section{Methodology}
\label{sec:Methodology}
\subsection{Task definition}
\begin{figure*}[htbp]
\centering
\includegraphics[width=\linewidth]{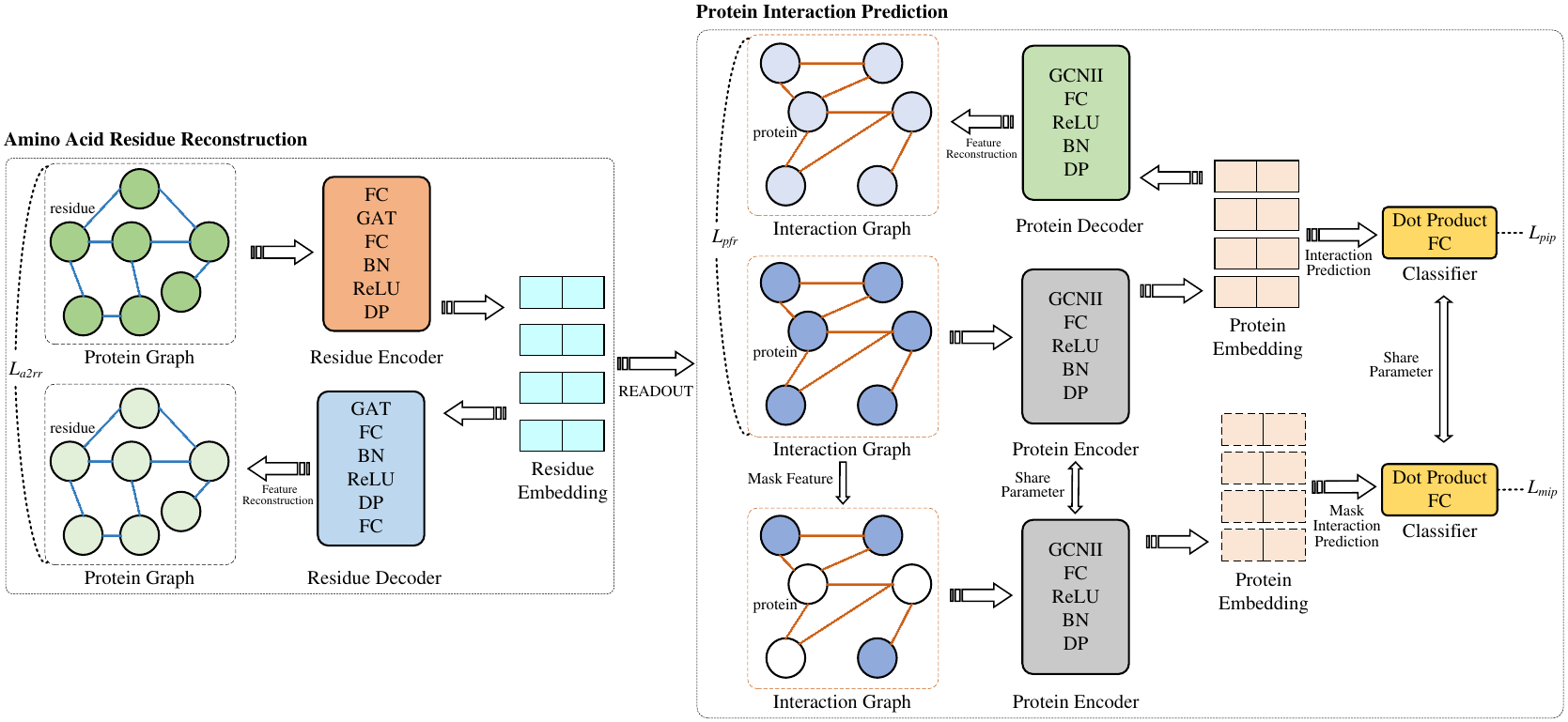}
\caption{Overall workflow of the proposed MgslaPPI.}
\label{fig:overall_workflow}
\end{figure*}
PPI prediction involves identifying the potential types of interactions between pairs of proteins based on their sequence, structure, and function characteristics. Mathematically, this can be described as: assuming there is a set of proteins \(Prot=\{prot_1, prot_2, \ldots, prot_M\}\), the task of PPI prediction is to infer the corresponding interaction type \(y_{ij}\) for a protein pair \(pair_{ij}=(prot_i, prot_j)\). As shown in Figure~\ref{fig:overall_workflow}, the MgslaPPI method proposed in this study consists of two stages, i.e., amino acid residue reconstruction (A2RR) and protein interaction prediction (PIP). Here, A2RR can be seen as a pre-training stage, while PIP is the core stage (or considered as the main stage) for PPI prediction. It should be noted that the input for A2RR is a protein graph, whereas the input for PIP is an interaction graph.

\textit{Protein Graph:} A protein $prot$ can be composed of multiple residues, and each residue can be described by various physicochemical properties. Given this, the protein graph is represented as \(G_{prot} = (V_{prot}, E_{prot})\), where a node \(v_i \in V_{prot}\) represents a residue in the protein \(prot\), and an edge \(e_{ij} \in E_{prot}\) represents the connection between residues \( v_i\) and \( v_j\). Following the mmanner of Gao et al.~\cite[]{gao2023hierarchical}, we consider 7 different amino acid residue features, such as polarity, acidity and alkalinity, hydrogen bond acceptor, etc.; according to the way of Zhang et al.~\cite[]{zhang2023protein}, we add edges between residue nodes with sequential distance 2.

\textit{Interaction Graph:} According to the structure of PPI network, we represent the interaction graph as \( G_{ppi} = (N_{ppi}, R_{ppi}) \). Here, a node \( n_i \in N_{ppi} \) represents a protein in the PPI network (i.e., \( prot_i \)), and an edge \( r_{ij} \in R_{ppi} \) represents the relationship between proteins \( n_i \) and \( n_j \); in MgslaPPI, we take the mean of all residues in a protein as the representation of that protein. Therefore, the goal of PPI prediction can be transformed into learning a function \( F \) that predicts the corresponding interaction types \( Y \) based on the interaction graph \( G_{ppi} \).

\subsection{Amino acid residue reconstruction}
We employ a residue reconstruction task based on GAT to capture the structural features of proteins. First, the constructed protein graph \( G_{prot} \) is input into the GAT module to mine the dependency information between amino acid residues. Then, the obtained residue representations are sequentially passed through a fully connected layer, a ReLU function, a batch normalization layer, and a dropout layer to enhance the feature expression of the residues. Here, we refer to the aforementioned networks as the residue encoder. After stacking multiple layers of the network, the residue representations extracted by this encoder can be obtained. Mathematically, the above process can be represented as:
\begin{equation}
\label{eq:residue_encoder}
\begin{split}
&X_f = \mathtt{FC}(X), \\
&X_g = \mathtt{GAT}(G_{prot}, X_f), \\
&\hat{X} = \mathtt{DP}(\mathtt{BN}(\mathtt{ReLU}(\mathtt{FC}(X_g)))),
\end{split}
\end{equation}
where \(\mathtt{FC}\), \(\mathtt{BN}\), and \(\mathtt{DP}\) represent the fully connected, batch normalization, and dropout layers, respectively; \(\mathtt{GAT}\) denotes the graph attention layer, which can be further mathematized as:
\begin{equation}
\begin{split}
&x_{g,i} = \sum _{v_j \in \mathtt{N}(v_i)} \alpha _{ij} W_g x_j, \\
&\mathtt{s.t.}\ \alpha _{ij} = \frac{\exp (\sigma(\mathrm{a}^\top [W x_i | | W x_j]))}
     {\sum_{v_k \in \mathtt{N}(v_i)} \exp (\sigma(\mathrm{a}^\top [W x_i | | W x_k]))},
\end{split}
\end{equation}
where \(\mathtt{N}(v_i)\) represents the neighbors of node (or residue) \(v_i\); \(W_g\), \(\mathrm{a}\), and \(W\) represent trainable parameters, and \(\sigma\) represents the LeakyReLU activation function. Following Veličković et al.~\cite[]{veličković2018graph}, we employ multi-head GAT to enhance the stability of the network.

In the residue decoding stage, we use the symmetric structure of the residue encoder, i.e., equation (\ref{eq:residue_encoder}), as the residue decoder to output the reconstructed features \(X^{\prime}\) corresponding to the original residue features \(X\). This stage can be mathematically represented as:
\begin{equation}
\begin{split}
&\hat{X}_g = \mathtt{GAT}(G_{prot}, \hat{X}), \\
&X_d = \mathtt{DP}(\mathtt{BN}(\mathtt{ReLU}(\mathtt{FC}(\hat{X}_g)))), \\
&X^{\prime} = \mathtt{FC}(X_d).
\end{split}
\end{equation}

For protein \( n_m \) (also denoted as \( prot_m \)), to make the reconstructed residue features \( X_m^{\prime} \) as close as possible to the original residue features \( X_m \), we define a training objective:
\begin{equation}
L_{a2rr}=\frac{1}{U} \sum_{u=1}^{U} (x_{m,u} - x_{m,u}^{\prime})^2,
\end{equation}
where \( U \) represents the total number of residues in protein \( n_m \), and \( x_{m,u} \in X_m \) represents the representation of the \( u \)-th residue in \( n_m \).

After training for residue reconstruction, a residue encoder capable of capturing the structural features of proteins is obtained. Finally, the representation of the protein can be obtained by performing a mean operation on the output of this encoder:
\begin{equation}
h_m = \mathtt{MEAN}(\hat{X}_{m}), \quad m = 1, 2, \ldots, M,
\end{equation}
where \( M \) denotes the total number of proteins, and \( h_m \in H \) is the representation of the \( m \)-th protein.

\subsection{Protein interaction prediction}
Prior studies~\cite[]{gao2023hierarchical,ijcai2021p506,Zeng2024GNNGL-PPI,10.1093/bioinformatics/btad052} have demonstrated that GNNs are effective in capturing interaction information between proteins. Building on this, we employ GCNs to model protein interactions. Specifically, we utilize GCNII~\cite[]{pmlr-v119-chen20v} as the main component of our encoder, with the constructed interaction graph \( G_{ppi} \) serving as the input. After stacking multiple layers, the GCNII model can thoroughly mine the associative information between proteins. This process can be mathematically formulated as:
\begin{equation}
H_g^{(l)} = \mathtt{GCNII}(G_{ppi}, H^{(l-1)}),
\end{equation}
where \( H^{(0)} \) represents the initial features of the proteins, i.e., \( H^{(0)} = [h_1, h_2, \ldots, h_M] \). The \(\mathtt{GCNII}\) can be further represented as:
\begin{equation}
\begin{split}
H_g^{(l)} = &((1-\beta_{l-1}) \tilde{P} H^{(l-1)} + \beta_{l-1} H^{(0)}) ((1-\gamma_{l-1}) I \\
&+ \gamma_{l-1} W^{(l-1)}),
\end{split}
\end{equation}
where \(\tilde{P} = \tilde{D}^{-1/2} \tilde{A} \tilde{D}^{-1/2}\), where \(\tilde{A}\) denotes the adjacency matrix with self-loops, and \(\tilde{D}_{ii} = \sum_{j=1} \tilde{A}_{ij}\); \(\beta_{l-1}\) and \(\gamma_{l-1}\) are hyperparameters, \(I\) represents the identity matrix, and \(W^{(l-1)}\) is a learnable parameter.

To enhance the representation of proteins, we pass the obtained protein representations through a fully connected layer, a ReLU function, a batch normalization layer, and a dropout layer. Mathematically, this can be expressed as:
\begin{equation}
\begin{split}
&\check{H} = \mathtt{DP}(\mathtt{BN}(\mathtt{ReLU}(\mathtt{FC}(H_g)))), \\
&\hat{H} = \mathtt{DP}(\mathtt{ReLU}(\mathtt{FC}(\check{H}))).
\end{split}
\end{equation}
Note that we consider the above networks as a whole, i.e., the protein encoder. To perform PPI prediction, we integrate the features of proteins \( n_i \) and \( n_j \) through a dot product operation and then feed them into a fully connected layer. Mathematically, this can be represented as:
\begin{equation}
\hat{y}_{ij} = FC(\hat{h}_i \cdot \hat{h}_j).
\end{equation}

Given the training set \( D_{train} = (\mathbb{X}, \mathbb{Y}) \), we use cross-entropy as the training objective to train the protein encoder. Mathematically, this can be expressed as:
\begin{equation}
\begin{split}
L_{pip} = &\frac{1}{|\mathbb{X}| \times C} \sum_{(n_i,n_j) \in \mathbb{X}} \sum_{c=1}^{C} \left(y_{ij}^c \log(\hat{y}_{ij}^c) \right. \\
&\left. + (1-y_{ij}^c) \log(1-\hat{y}_{ij}^c) \right),
\end{split}
\end{equation}
where \(\mathbb{X}\) represents the protein pairs in the training set, \(\mathbb{Y}\) represents the corresponding ground-truth interaction types, and \(C\) represents the total number of PPI types. 

\subsection{Auxiliary task}
Relevant studies~\cite[]{10.1007/978-3-031-72989-8_14,yang2024adamerging,10502283,10735791} have shown that introducing auxiliary tasks can enhance the learning performance of the main task. Therefore, in the PIP stage, in addition to the interaction prediction task (i.e., main task), we have introduced two auxiliary tasks: protein feature reconstruction (PFR) and masked interaction prediction (MIP).

For the PFR task, we add a symmetric structure of the protein encoder to decode the protein features. The specific process can be mathematically represented as:
\begin{equation}
\begin{split}
&\hat{H}_g = \mathtt{GCNII}(G_{ppi}, \hat{H}), \\
&\check{H}^{\prime} = \mathtt{DP}(\mathtt{BN}(\mathtt{ReLU}(\mathtt{FC}(\hat{H}_g)))), \\
&H^{\prime} = \mathtt{DP}(\mathtt{ReLU}(\mathtt{FC}(\check{H}^{\prime}))).
\end{split}
\end{equation}

To make the output of the protein decoder \( H' \) (i.e., the reconstructed protein features) as close as possible to the original protein features \( H \), we define a protein reconstruction objective:
\begin{equation}
L_{pfr} = \frac{1}{M} \sum_{m=1}^{M} (h_m - h_m^{\prime})^2,
\end{equation}
where \( M \) represents the total number of proteins, and \( h_m \in H \) represents the representation of the \( m \)-th protein.

Masked modeling has been widely applied in image analysis~\cite[]{Xie_2022_CVPR}, video generation~\cite[]{Yu_2023_CVPR}, language understanding~\cite[]{czinczoll-etal-2024-nextlevelbert}, and multimodal learning~\cite[]{NEURIPS2023_b6446566}. In the designed MIP task, we randomly mask protein features at a certain ratio and then pass the masked features through a shared protein encoder, and finally perform the mask interaction prediction. The mathematical expression is as follows:
\begin{equation}
\begin{split}
&H_{mask} = \mathtt{MASK}(H),\\
&\ddot{H}_g = \mathtt{GCNII}(G_{ppi}, H_{mask}), \\
&\ddot{H}_f = \mathtt{DP}(\mathtt{BN}(\mathtt{ReLU}(\mathtt{FC}(\ddot{H}_g)))), \\
&\ddot{H} = \mathtt{DP}(\mathtt{ReLU}(\mathtt{FC}(\ddot{H}_f))),\\
&\ddot{y}_{ij} = \mathtt{FC}(\ddot{h}_i \cdot \ddot{h}_j).
\end{split}
\end{equation}

Similar to conventional interaction prediction, we use cross-entropy as the training objective:
\begin{equation}
\begin{split}
L_{mip} = &\frac{1}{|\mathbb{X}| \times C} \sum_{(n_i,n_j) \in \mathbb{X}} \sum_{c=1}^{C} \left(y_{ij}^c \log(\ddot{y}_{ij}^c) \right. \\
&\left. + (1-y_{ij}^c) \log(1-\ddot{y}_{ij}^c) \right).
\end{split}
\end{equation}

In the PIP stage, we have a total of three tasks, including one main task (i.e., conventional interaction prediction) and two auxiliary tasks (i.e., protein feature reconstruction and masked interaction prediction). Combining these three tasks:
\begin{equation}
L = L_{pip} + L_{pfr} + L_{mip} + \eta  \lvert W_l \rvert,
\end{equation}
where \(\eta\) represents the L2 regularization weight, and \(W_l\) is a learnable parameter.

\section{Experiment}
\label{sec:Experiment}
\subsection{Dataset}
In this study, we evaluate the proposed method on the SHS27K and SHS148K datasets, following the way of Chen et al.~\cite[]{10.1093/bioinformatics/btz328}. The SHS27K and SHS148K datasets are subsets of human PPI derived from the STRING database~\cite[]{10.1093/nar/gky1131}. In accordance with the method of Chen et al., we randomly select proteins from the STRING database with more than 50 amino acids and sequence homology below 40\%, resulting in the SHS27K and SHS148K datasets containing 16,912 and 99,782 PPI data, respectively. These datasets encompass seven different types of PPIs, namely reaction, binding, post-translational modifications (ptmod), activation, inhibition, catalysis, and expression. Random partitioning methods can often lead to significant overlap between proteins in the training and test sets, thereby causing overfitting and inflated performance metrics on the respective datasets. To maintain consistency with previous works~\cite[]{ijcai2021p506,wu2024mapeppi}, we construct the test sets using breadth-first search (BFS), depth-first search (DFS), and random partitioning schemes. Following the partitioning scheme of Wu et al.~\cite[]{wu2024mapeppi}, we divide the two datasets into training, validation, and test sets in a 3:1:1 ratio. The detailed data division is shown in Table~\ref{tab:dataset}
\begin{table*}[htbp]
\centering
\scriptsize
\setlength{\tabcolsep}{12pt}
\caption{\label{tab:dataset}Data partitioning for two datasets.}
\begin{tabular}{cccccccccc}
\toprule
\multirow{2}{*}{Dataset}  & \multicolumn{3}{c}{Random} &\multicolumn{3}{c}{BFS} &\multicolumn{3}{c}{DFS}\\
\cmidrule(lr){2-4} \cmidrule(lr){5-7} \cmidrule(lr){8-10} 
     & training & validation & test & training & validation & test & training & validation & test \\
\midrule
SHS27K &4,440 &1,480 & 1,481 &4,427 &1,480 &1,494 &4,427 &1,480 &1,494\\
SHS148K &26,038 &8,679 &8,680 &26,000 &8,679 &8,718 & 26,018 &8,679 &8,700 \\
\bottomrule
\end{tabular}
\end{table*}

\subsection{Baseline}
To demonstrate the effectiveness of our MgslaPPI, we select a variety of baseline models for comparison. These baseline models are divided into two major categories: sequence-based methods and structure-based methods. The sequence-based methods include DPPI~\cite[]{10.1093/bioinformatics/bty573}, DNN-PPI~\cite[]{molecules23081923}, PIPR~\cite[]{10.1093/bioinformatics/btz328}, GNN-PPI~\cite[]{ijcai2021p506}, and SemiGNN-PPI~\cite[]{zhao2023semignn}. The structure-based methods comprise HIGH-PPI~\cite[]{gao2023hierarchical} and MAPE-PPI~\cite[]{wu2024mapeppi}.

DPPI models and predicts PPI based on sequence information, leveraging deep learning techniques to not only efficiently process large amounts of training data but also capture complex relationships involved in interactions. 
DNN-PPI is a deep neural network framework that sequentially inputs raw amino acid sequences into encoding, embedding, CNN, and LSTM layers, automatically learning features directly from the primary sequence of proteins to predict PPI.
PIPR is an end-to-end framework based on recurrent neural network, with an architecture based on residual recurrent convolutional neural networks, effectively integrating local features and contextual information to predict protein interactions.
GNN-PPI designs a new data partitioning scheme for the interactions of novel proteins, employing breadth-first and depth-first search methods to construct test sets in addition to randomizing the data.
SemiGNN-PPI constructs and processes multiple graphs from the perspective of features and labels, exploring label dependencies, and combines GNN with Mean Teacher to effectively utilize unlabeled graph-structured PPI data.
HIGH-PPI is a dual-view hierarchical graph learning model. In the bottom view, HIGH-PPI constructs protein graphs with amino acid residues as nodes and physical adjacency as edges; in the top view, it constructs PPI graphs with protein graphs as nodes and interactions as edges, employing TGNN to learn protein-protein relationships.
MAPE-PPI combines the chemical properties and geometric features surrounding amino acid residues, proposing microenvironment-aware protein embeddings, and introduces a novel masked codebook modeling strategy for pretraining data to capture dependencies between different microenvironments.

\subsection{Implementation detail}
The operating system used is Ubuntu 20.04, and the deep learning framework is Pytorch 2.4.1. All experiments are conducted on a device equipped with a single NVIDIA GeForce RTX 3090 GPU. We adopt AlphaFold2~\cite[]{Jumper2021Highly} to retrieve the 3D structure of the amino acid sequence data. In the experiments, we employ the Adam optimizer~\cite[]{KingmaB14Adam} with an L2 regularization coefficient of 1e-4 and a learning rate of 1e-3. During the A2RR phase, both the encoder and decoder have 4 network layers, the number of heads for GAT is set to 8, the hidden layer dimension is configured to 128, the batch size was 128, and the maximum number of epochs is set to 50. In the PIP phase, the number of network layers for GCNII is adjusted to 3, the hidden layer dimension is 1024, the masking rate is set to 20\%, and the maximum number of epochs is 1000. Considering the class imbalance issue in the dataset, following previous studies (such as GNN-PPI~\cite[]{ijcai2021p506}, SemiGNN-PPI~\cite[]{zhao2023semignn}, MAPE-PPI~\cite[]{wu2024mapeppi}, etc.), we adopt the micro-F1 score as the evaluation metric for this work.

\subsection{Main result}
\begin{table*}[htbp]
\centering
\scriptsize
\setlength{\tabcolsep}{9.5pt}
\caption{\label{tab:main_result}Main results on two datasets. All baseline models' scores are derived from the experimental results of Wu et al.~\cite[]{wu2024mapeppi}, with bold and underlining indicating the best and second-best performances, respectively.}
\begin{tabular}{cccccccccc}
\toprule
\multirow{2}{*}{Method}  & \multicolumn{4}{c}{SHS27K} &\multicolumn{4}{c}{SHS148K} & \multirow{2}{*}{Average}\\
\cmidrule(lr){2-5}  \cmidrule(lr){6-9} 
 & Random & BFS & DFS & Average & Random & BFS & DFS & Average & \\
\midrule
DPPI~\cite[]{10.1093/bioinformatics/bty573} &70.45 &43.87 & 43.69 &52.67 &76.10 &50.80 & 51.43 &59.44 & 56.06\\
DNN-PPI~\cite[]{molecules23081923} &75.18 &51.59 &48.90 & 58.56 &85.44 &54.56 & 56.70 &65.57 & 62.06\\
PIPR~\cite[]{10.1093/bioinformatics/btz328} & 79.59 &47.13 &52.19 & 59.64 &88.81 &58.57 & 61.38 &69.59 & 64.61\\
GNN-PPI~\cite[]{ijcai2021p506} &83.65 &63.08 &66.52 &71.08 & 90.87 &69.53 &75.34 &78.58 & 74.83\\
SemiGNN-PPI~\cite[]{zhao2023semignn} &85.57 &67.94 &69.25 &74.25 & 91.40 &71.06 &77.62 &80.03 & 77.14\\
HIGH-PPI~\cite[]{gao2023hierarchical} &86.23 &68.40 &70.24 &74.96 & 91.26 &72.87 &78.18 &80.77 & 77.86\\
MAPE-PPI~\cite[]{wu2024mapeppi} &\textbf{88.91} &\underline{70.38} &\underline{71.98} &\underline{77.09} & \underline{92.38} &\underline{74.76} &\underline{79.45} &\underline{82.20} & \underline{79.64}\\
\midrule
MgslaPPI (Ours) & \underline{88.61} & \textbf{78.13} & \textbf{73.10} &\textbf{79.95} & \textbf{92.57} & \textbf{76.49} & \textbf{82.29} &\textbf{83.78} & \textbf{81.87}\\
\bottomrule
\end{tabular}
\end{table*}
In Table~\ref{tab:main_result}, we present the comparative results of MgslaPPI and baseline methods on two datasets, covering three different data partitioning methods. Among all baseline models, HIGH-PPI and MAPE-PPI rank second and first in performance, respectively, while the traditional GNN-PPI shows obvious limitations in terms of F1 score. Our MgslaPPI outperforms all baseline methods in average F1 score, achieving a 2.23\% improvement over the best-performing baseline, MAPE-PPI. Under two more challenging data partitioning schemes (BFS and DFS)~\cite[]{ijcai2021p506}, our method realizes significant performance enhancements. For instance, under the DFS partitioning scheme, MgslaPPI's F1 scores on the SHS27K and SHS148K datasets are 1.12\% and 2.84\% higher than those of MAPE-PPI, respectively. Notably, the most significant improvement of MgslaPPI occurs under the BFS partitioning scheme of the SHS27K dataset, where it outperforms MAPE-PPI by 7.75\%. These performance improvements indicate that our method can effectively mine the internal structural information and external interaction information of proteins, thereby enhancing the accuracy of PPI prediction. It is worth mentioning that we found all sequence-based baseline methods to be inferior in performance to structure-based methods, including HIGH-PPI, MAPE-PPI, and MgslaPPI. This phenomenon suggests that the internal structural information of proteins can enrich the expression of proteins themselves, and capturing this information can further improve the performance of PPI prediction.

\subsection{Ablation study}
\begin{table*}[htbp]
\centering
\scriptsize
\setlength{\tabcolsep}{11.7pt}
\caption{\label{tab:ablation_study}Results of ablation studies on two datasets.}
\begin{tabular}{cccccccccc}
\toprule
\multirow{2}{*}{Method}  & \multicolumn{4}{c}{SHS27K} &\multicolumn{4}{c}{SHS148K} & \multirow{2}{*}{Average}\\
\cmidrule(lr){2-5}  \cmidrule(lr){6-9} 
      & Random & BFS & DFS & Average & Random & BFS & DFS & Average & \\
\midrule
MgslaPPI & \textbf{88.61} & \textbf{78.13} & \textbf{73.10} &\textbf{79.95} & \textbf{92.57} & \textbf{76.49} & \textbf{82.29} &\textbf{83.78} & \textbf{81.87}\\
\midrule
w/o GAT & 75.03  & 69.26  & 67.69 & 70.66 & 82.83 & 63.97 & 73.36 & 73.39 & 72.02\\
w/o GAT w GCN & 88.11 & 76.43 & 72.54 & 79.03 & 92.48 & 76.30 & 80.88 & 83.22 & 81.12 \\
\midrule
w/o $L_{pfr}$ & 88.13 & 77.11 & 70.67 &78.64 & 91.75 & 73.99 & 80.84 &82.19 &80.42 \\
w/o $L_{mip}$ & 88.20 & 74.78 & 71.48 &78.15 & 92.24 & 74.73 & 81.68 &82.88 &80.52 \\
w/o $L_{pfr}$ w/o $L_{mip}$ & 88.14 & 76.13 & 72.06 &78.78 & 92.14 & 73.42 & 81.09 &82.22 &80.50 \\
\bottomrule
\end{tabular}
\end{table*}
In this subsection, we conduct a series of ablation studies to examine the effectiveness of each component of MgslaPPI, with the experimental results shown in Table~\ref{tab:ablation_study}. We perform ablation experiments on the GAT module in the A2RR stage in two ways: one is to directly remove the GAT module and obtain protein representations using directly residue features; the other is to replace the GAT module with a GCN module~\cite[]{kipf2017semisupervised}. The experimental results indicate that after removing the GAT module, there is a significant performance decline in all data partitions of the two datasets. Specifically, in the SHS27K dataset, the performance decreases by 13.58\%, 8.87\%, and 5.41\% under the Random, BFS, and DFS partitioning schemes, respectively; in the SHS148K dataset, the performance decreases by 9.74\%, 12.52\%, and 8.93\% under the three partitioning schemes. When the GCN module is used to replace the GAT module, the experimental results also show a slight decline, with the average F1 scores of the two datasets decreasing by 0.92\% and 0.56\%, respectively. These phenomena suggest that the GAT module in MgslaPPI can effectively capture the structural information of proteins and has a positive impact on PPI prediction.

To verify the effectiveness of the designed auxiliary tasks in improving model performance, we carry out three types of experiments: one is to remove the protein feature reconstruction (PFR) task; the second is to eliminate the masked interaction prediction (MIP) task; and the third is to remove both of the above tasks. The experimental results show that after removing the PFR task, the model performance decreases in all three data partitioning schemes of the two datasets. For example, in the BFS partitioning of the SHS148K dataset, the F1 score decreases by 2.5\%, indicating that the PFR task helps the protein encoder capture more useful information. Similarly, when the MIP task is removed, a decrease in the experimental results is also observed. For instance, in the BFS partitioning of the SHS27K dataset, the performance decreases by 3.35\%, proving that the MIP task can promote the main task to fully explore the interaction information between proteins. In addition, when both the PFR and MIP tasks are removed, the F1 scores of the model on the two datasets also decrease.

\subsection{In-depth analysis of different subsets}
\begin{figure*}[htbp]
\centering
{\includegraphics[width=0.3\linewidth]{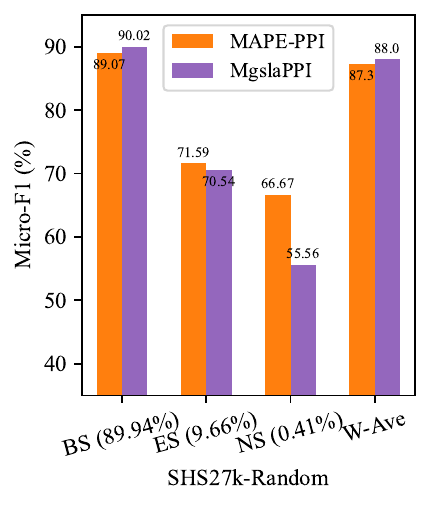}%
\label{fig:in-depth_analysis_SHS27k-Random}}
\hfil
{\includegraphics[width=0.3\linewidth]{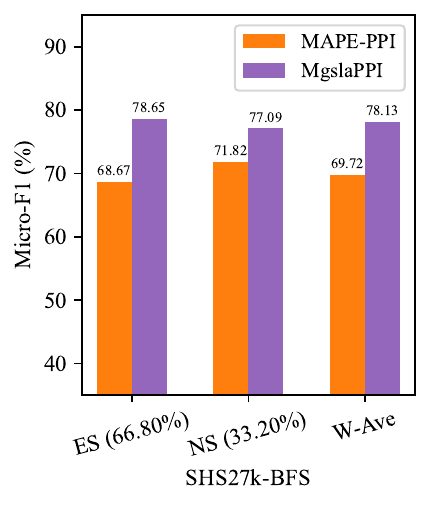}%
\label{fig:in-depth_analysis_SHS27k-BFS}}
\hfil
{\includegraphics[width=0.3\linewidth]{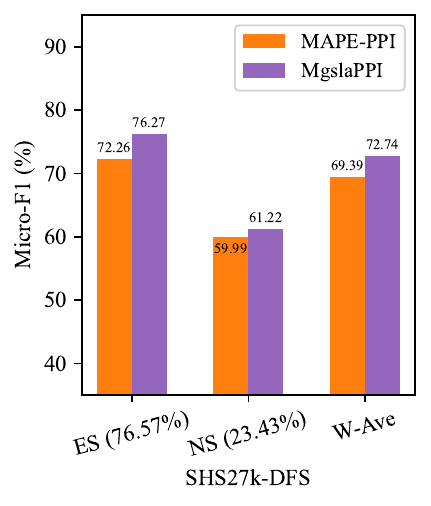}%
\label{fig:in-depth_analysis_SHS27k-DFS}}
\caption{Results on different subsets (BS, ES, and NS) of the SHS27k dataset. F1 scores of MAPE-PPI are reproduced based on the official code~\cite[]{wu2024mapeppi}, and W-Ave denotes the weighted average.}
\label{fig:in-depth_analysis}
\end{figure*}
To conduct an in-depth analysis, we divide the PPIs in the test set into three subsets: the BS (both proteins exist in the labeled training set), the ES (at least one protein exists in the labeled training set), and the NS (neither protein exists in the labeled training set) subsets. As can be seen from Figure~\ref{fig:in-depth_analysis}, the BFS and DFS partitions only include the ES and NS subsets, while the BS subset accounts for the vast majority in the Random partition. This phenomenon indicates that the BFS and DFS partitions are more challenging than the Random partition. From the experimental results, it can be observed that MgslaPPI outperforms MAPE-PPI in these two more challenging data partitioning schemes. For example, on the ES subsets of BFS and DFS partitions, the F1 scores of MgslaPPI are 9.98\% and 4.01\% higher than those of MAPE-PPI, respectively; on the NS subsets of the two partitions, the differences between the two methods are 5.27\% and 1.23\%, respectively. Looking at the weighted average scores of the three partitions, MgslaPPI is also higher than MAPE-PPI. From the above data, it can be concluded that our PPI prediction method performs excellently on unseen datasets.

\subsection{Result for each category}
\begin{figure*}[htbp]
\centering
{\includegraphics[width=0.3\linewidth]{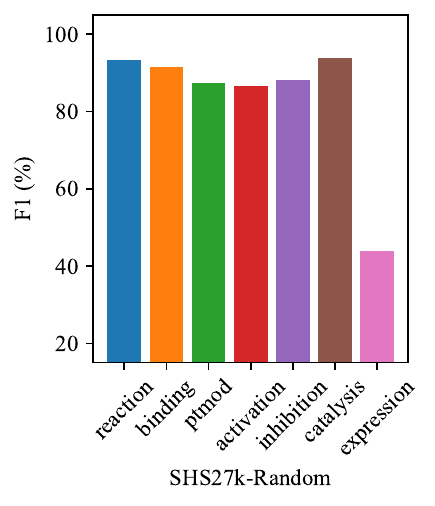}%
\label{fig:each_category_SHS27k-Random}}
\hfil
{\includegraphics[width=0.3\linewidth]{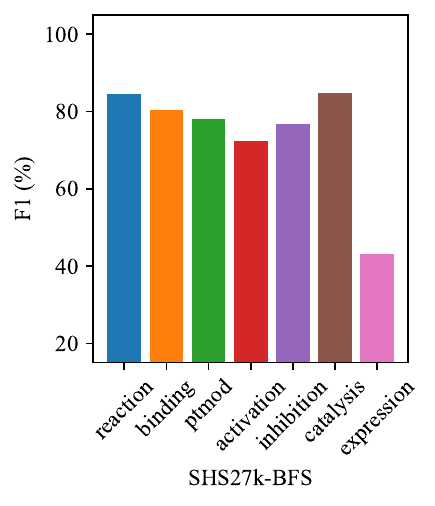}%
\label{fig:each_category_SHS27k-BFS}}
\hfil
{\includegraphics[width=0.3\linewidth]{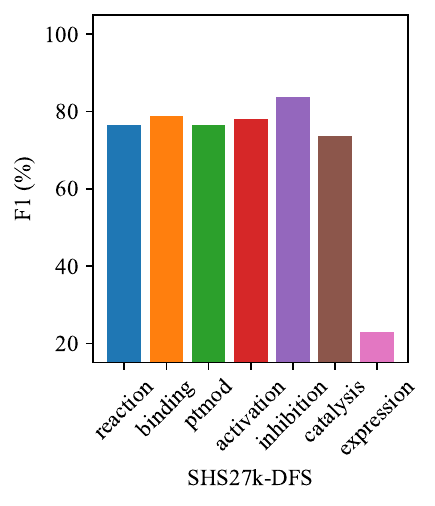}%
\label{fig:each_category_SHS27k-DFS}}
\caption{Accuracy for each category on the SHS27k dataset.}
\label{fig:each_category}
\end{figure*}
Figure~\ref{fig:each_category} presents the accuracy of each PPI type on the SHS27k dataset under the three partitioning schemes. Among all partitioning schemes, the \textit{expression} type is more difficult for MgslaPPI to recognize compared to other PPI types. Upon examining the distribution of the SHS27k dataset, it is found that there is a class imbalance issue, with the \textit{expression} type being a minority class. Therefore, the low accuracy of \textit{expression} may be due to the model's tendency to learn the majority classes. Compared to Random and BFS partitioning, the accuracy of \textit{expression} under DFS partitioning is even lower. Further inspection of the data reveals that the number of \textit{expression} instances under the DFS partitioning scheme is less than that under the other two partitioning schemes. Overall, it can be seen that each PPI type is easiest to classify correctly under Random partitioning, while the accuracy of most PPI types (except for \textit{activation} and \textit{inhibition}) is the lowest under DFS partitioning, indicating that the latter is more challenging.

\subsection{Impact of different mask rates}
\begin{figure*}[htbp]
\centering
{\includegraphics[width=0.45\linewidth]{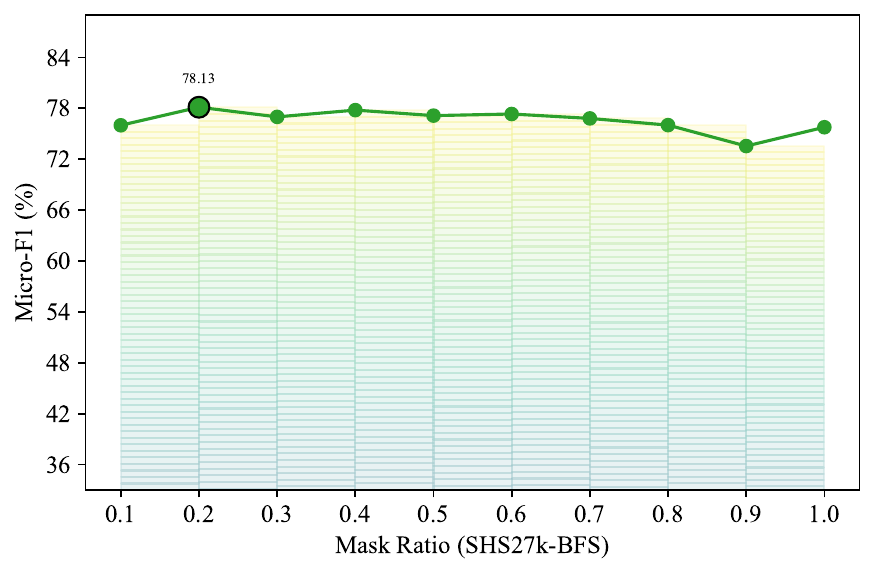}%
\label{fig:mask2_ratio_SHS27k-BFS}}
\hfil
{\includegraphics[width=0.45\linewidth]{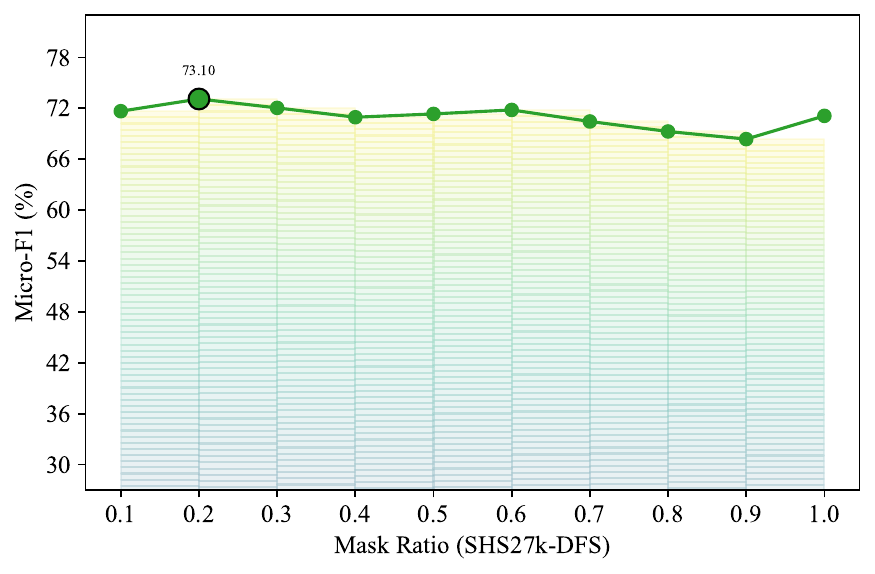}%
\label{fig:mask2_ratio_SHS27k-DFS}}
\caption{Performance variation with network depth on the SHS27k dataset.}
\label{fig:mask2_ratio}
\end{figure*}
As shown in Figure~\ref{fig:mask2_ratio}, we conduct experiments with different masking rates on the SHS27k dataset under both BFS and DFS partitioning methods, and record the corresponding F1 scores. This experiment aims to investigate the impact of different masking rates on the performance of MgslaPPI in the MIP task. The results indicate that when the masking rate is set to 0.2, the performance of MgslaPPI is optimal under both BFS and DFS partitions. As the masking rate is adjusted, the experimental results of MgslaPPI fluctuate around the best score.

\subsection{Impact of different network depths}
\begin{figure*}[htbp]
\centering
{\includegraphics[width=0.45\linewidth]{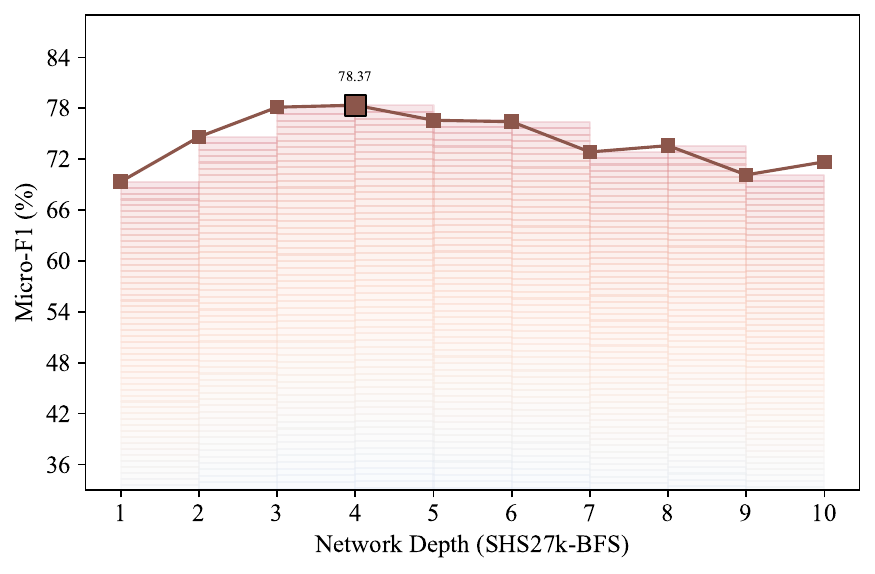}%
\label{fig:network_depth_SHS27k-BFS}}
\hfil
{\includegraphics[width=0.45\linewidth]{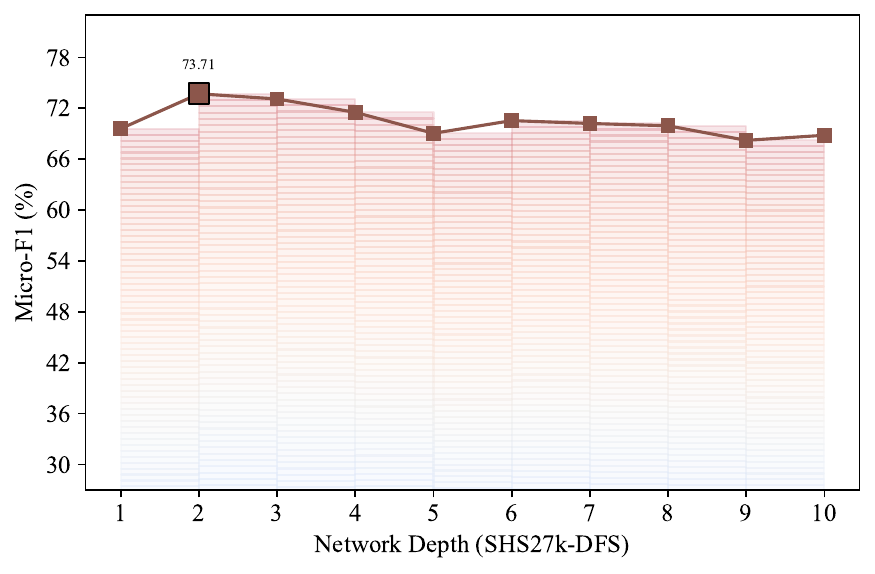}%
\label{fig:network_depth_SHS27k-DFS}}
\caption{Performance variation with network depth on the SHS27k dataset.}
\label{fig:network_depth}
\end{figure*}
To investigate the impact of different network depths on the performance of MgslaPPI, we have presented in Figure~\ref{fig:network_depth} the performance changes of the SHS27k dataset under two partitioning manners, BFS and DFS. It can be observed from the figure that under BFS partitioning, the best performance is achieved when the network depth is 4; while under DFS partitioning, the optimal result is obtained when the number of network layers is 2. This indicates that for different data partitioning manners, the network depth corresponding to the best F1 performance varies, and this pattern can also be generalized to different datasets. Overall, regardless of the partitioning manner used, as the network depth increases, the F1 score of the model first rises and then falls after reaching a certain peak value.

\section{Conclusion}
This paper addresses the challenges of PPI prediction by proposing a novel method named MgslaPPI based on multitask graph structure learning. This method decomposes PPI prediction into two stages: amino acid residue reconstruction (A2RR) and protein interaction prediction (PIP), leveraging GNNs to respectively extract the internal structure and external interaction information of proteins. In the A2RR stage, GAT is utilized to model proteins, capturing the dependencies among amino acid residues to obtain the internal structural information of proteins. In the PIP stage, GCN is employed as encoders to model the PPI network, extracting interaction information between proteins, and two auxiliary tasks—protein feature reconstruction and masked interaction prediction—are introduced to enhance the expressive power of the graph encoder and improve the expression of protein information. Through comparison and ablation extensive experiments on two public PPI datasets, the results demonstrate that MgslaPPI outperforms existing advanced prediction methods under multiple data partitions.








\balance
\bibliographystyle{elsarticle-harv}
\bibliography{mgslappi.bib}

\end{document}